\documentclass[10pt, conference]{IEEEtran}

\IEEEoverridecommandlockouts
\usepackage{cite}
\usepackage{amsmath,amssymb,amsfonts}
\usepackage{algorithmic}
\usepackage{graphicx}
\usepackage{textcomp}
\usepackage{xcolor}
\usepackage{subcaption} 
\usepackage{array}
\def\BibTeX{{\rm B\kern-.05em{\sc i\kern-.025em b}\kern-.08em
    T\kern-.1667em\lower.7ex\hbox{E}\kern-.125emX}}
\begin{document}

\begin{titlepage}
\thispagestyle{empty}
\centering

\vspace*{\fill}

{\Huge\bfseries Preprint\par}
\vspace{1cm}

{\Large Accepted paper at IEEE HPSR 2026\par}
\vspace{2.5cm}

\begin{center}
\begin{minipage}{0.6\textwidth}
\centering
\small
\textcopyright\ 2026 IEEE. Personal use of this material is permitted.
Permission from IEEE must be obtained for all other uses, in any current
or future media, including reprinting/republishing this material for
advertising or promotional purposes, creating new collective works,
for resale or redistribution to servers or lists, or reuse of any
copyrighted component of this work in other works.
\end{minipage}
\end{center}

\vspace{1.5cm}

{\large Until published, please cite as:\par}

\vspace{0.5cm}

\begin{center}
\begin{minipage}{0.65\textwidth}
\centering
\small
Sarpkaya, F. B., Francini, A., Erman, B., Panwar, S.,
Evaluation of TCP Congestion Control for Public High-Performance Wide-Area Networks,
Proc. 27th International Conference on High Performance Switching and Routing,
IEEE HPSR 2026, Montreal, Canada, June 2026, in print.
\end{minipage}
\end{center}
\vspace*{\fill}

\end{titlepage}

\setcounter{page}{1}

\title{Evaluation of TCP Congestion Control for Public High-Performance Wide-Area Networks}

\author{
Fatih Berkay Sarpkaya$^{1}$\thanks{Fatih Berkay Sarpkaya was with Nokia Bell Labs, Network Systems and Security Research, Murray Hill, NJ (USA) when this work was performed.}, Andrea Francini$^{2}$, Bilgehan Erman$^{2}$, Shivendra Panwar$^{1}$\\[0.4em]
\small
$^{1}$\textit{NYU Tandon School of Engineering}, Brooklyn, NY, USA\\
$^{2}$\textit{Nokia Bell Labs}, Murray Hill, NJ, USA\\[0.3em]
\small
fbs6417@nyu.edu, andrea.francini@nokia-bell-labs.com, bilgehan.erman@nokia-bell-labs.com, panwar@nyu.edu
}

\maketitle

\begin{abstract}
Practitioners of a growing number of scientific and artificial-intelligence (AI) applications use High-Performance Wide-Area Networks (HP-WANs) for moving massive data sets between remote facilities.
Accurate prediction of the flow completion time (FCT) is essential in these data-transfer workflows because compute and storage resources are tightly scheduled and expensive. We assess the viability of three TCP congestion control algorithms (CUBIC, BBRv1, and BBRv3) for massive data transfers over {\em public\/} HP-WANs, where limited control of critical data-path parameters precludes the use of Remote Direct Memory Access (RDMA) over Converged Ethernet (RoCEv2), which is known to outperform TCP in {\em private\/} HP-WANs. 
Extensive experiments on the FABRIC testbed indicate that the configuration control limitations can also hinder TCP, especially through microburst-induced packet losses. Under these challenging conditions, we show that the highest FCT predictability is achieved by combination of BBRv1 with the application of traffic shaping before the HP-WAN entry points.

\end{abstract}

\section{Introduction}
\label{sec:introduction}
Large-scale scientific applications, in fields such as weather forecasting, molecular biology, and astrophysics, generate massive data sets that require high-performance computing (HPC) facilities. When these facilities are not available locally, the data must be transferred reliably over high-capacity, long-distance networks. A similar scenario arises in AI applications, for example in GPU-as-a-service (GPUaaS) arrangements, where users transfer massive amounts of data to and from HPC facilities before and after workload execution.

HPC reservations are expensive and typically scheduled in advance: high predictability of the data transfer time, or flow completion time (FCT), enhances the cost effectiveness of such reservations by reducing the need for safety margins~\cite{hpwan-framework}.
The maximum FCT value that a given configuration of the data transfer can yield is the defining metric for the file transfer contribution to the overall length of the HPC reservation.

We focus on FCT predictability in scenarios where the owner of the data transfer (the {\em user\/}) does not control the network infrastructure that supports it.  
The prevalence of such scenarios, where network-compute capacity is offered as a service to third-party users, is growing with the set of applications that require massive transfers between remote locations, extending the public-cloud business model.

Achieving predictable FCTs is a significant challenge when sharing a public high-performance wide-area network (HP-WAN) with massive data transfers from unrelated users. 
Accurate estimation of the FCT requires the capacity of the network path to be known and stable over time.
A virtual circuit (VC) with fixed, guaranteed bandwidth can provide the necessary traffic isolation through various combinations of traffic engineering and traffic management functions, including explicit routing, admission control, policing, shaping, flow control, and congestion control.  
Given the variety of possible combinations and the lack of user control over their configuration, the ideal behavior of the data transfer endpoints is one that achieves FCT predictability independently of the specific mechanisms used to enforce isolation.

Even with bandwidth isolation, predictable FCTs ultimately depend on end-to-end transport behavior. In practice, this necessitates a reliable transport such as TCP to ensure end-to-end delivery, and, critically, the selection of a congestion control (CC) algorithm that delivers consistent throughput under the large bandwidth-delay product (BDP) conditions of an HP-WAN VC.
Alternatives to TCP exist but are generally not better suited to the public HP-WAN use case that we are considering.
For example, the superior performance of Remote Direct Memory Access (RDMA) over Converged Ethernet (RoCEv2)~\cite{rocev2_specification} has been clearly proven in private HP-WAN setups~\cite{esnet100gbps}, but cannot be reproduced in the public case, where the user cannot fine-tune the network parameters for the mandatory lossless operation of RoCEv2~\cite{microsoft-rocev2}.
QUIC enables faster connection setup and easier customization of the CC algorithm than TCP, but in massive data transfers the setup time is practically negligible and a custom CC algorithm would only be warranted if existing TCP options proved inadequate. 
Finally, Multi-Path TCP (MPTCP) is out of scope, as it conflicts with our working assumption of a single end-to-end VC with fixed, guaranteed bandwidth.

Under generic network‑path assumptions, congestion control (CC) algorithms are typically evaluated based on fairness and adaptability to heterogeneous and time‑varying path conditions, including those in large bandwidth‑delay‑product (BDP) networks~\cite{fabric-largebdp}. In contrast, the bandwidth‑isolated VCs available in public HP‑WANs---e.g., realized via MPLS tunnels---allow us to abstract away fairness and adaptability concerns and focus instead on throughput predictability. Accordingly, we study the HP‑WAN viability of three representative and practically relevant TCP CC algorithms: CUBIC~\cite{cubic}, BBRv1~\cite{bbr2016}, and BBRv3~\cite{bbrv3-ietf-120-slides}. CUBIC serves as a loss‑based baseline, BBRv1 represents a model‑based design, and BBRv3 reflects the latest evolution of the BBR approach. Our evaluation is conducted on a multi‑site topology provisioned over the FABRIC research testbed~\cite{FABRIC}, leveraging inter‑site bandwidth‑isolation capabilities enabled by a recent testbed upgrade~\cite{fabric-qos}. Rather than benchmarking the full spectrum of TCP CC algorithms, our goal is to compare these representative designs in the specific HP‑WAN scenario under study. 
Our experiment artifact is publicly available on the FABRIC artifact portal~\cite{fabric_artifact_hpwan}.

\begin{figure}[ht]
    \centering
    \includegraphics[width=0.46\textwidth]{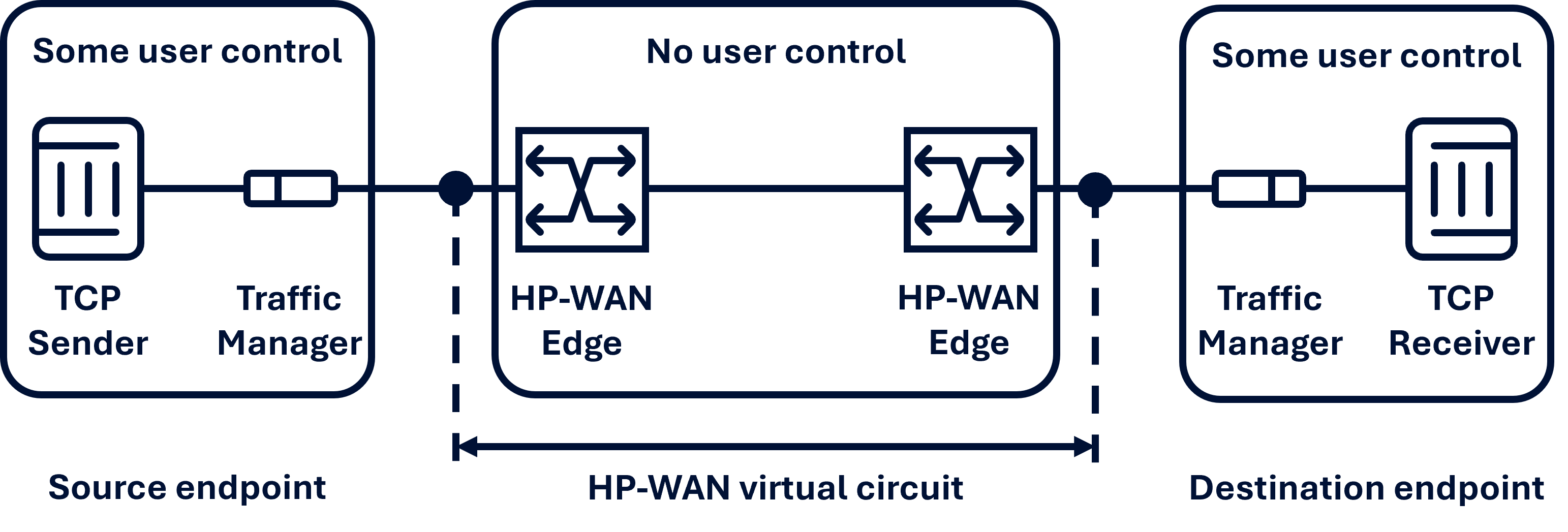}
    \caption{\small{Logical representation of the massive-transfer data path.}}
    \label{fig:representative-scenario}
\end{figure}

Fig.~\ref{fig:representative-scenario} shows the logical representation of the data path for massive data transfers between the endpoints.
The source and destination endpoints include traffic management devices that forward packets to and from the respective VC endpoints. 
In FABRIC, these devices are instantiated in software within {\em virtual machines\/} (VMs), with implementation options that include Linux traffic control tools and Data-Plane Development Kit (DPDK)~\cite{dpdk} applications.

We present an experimental evaluation of prominent TCP congestion control algorithms for massive transfers over public HP-WANs with bandwidth-isolated virtual circuits. Our main contributions are as follows:
\begin{itemize}

\item \textbf{We experimentally study TCP CC in public HP-WANs with bandwidth-isolated VCs.}
Using FABRIC Layer-2 Point-to-Point (L2PTP) VCs, we evaluate a setting where the WAN enforces bandwidth reservations, shifting the focus from fairness to FCT efficiency and predictability.

\item \textbf{We characterize the impact of non-congestive packet losses on TCP performance in the HP-WAN environment.}
Despite extensive tuning of host and data-path parameters, we observe packet losses uncorrelated with delay that can significantly degrade FCT predictability.

\item \textbf{We show that robustness to non-congestive losses is the dominant criterion for CC selection under bandwidth isolation.}
In this setting, BBRv1 achieves the most stable and predictable completion times, while CUBIC and BBRv3 are more adversely affected by such losses (CUBIC more than BBRv3).   

\item \textbf{We quantify the impact of the forwarding implementation at the VC ingress.}
Compared to Linux kernel forwarding, a DPDK implementation substantially improves FCT statistics by reducing delay jitter and bursty losses.

\item \textbf{We demonstrate that per-VC sender-side traffic shaping is the most effective endpoint configuration.}
Placing shapers upstream of the HP-WAN VC consistently yields high FCT efficiency and negligible retransmission overhead, outperforming alternative configurations. 

\item \textbf{We show that, with BBRv1, a single TCP connection fully utilizes VC bandwidths up to at least 20\,Gb/s.}
Striping across multiple TCP connections is unnecessary in this regime, simplifying the endpoint implementation and avoiding striping‑related overheads~\cite{hercules}.
    
\end{itemize}

The rest of this paper is organized as follows. 
Section~\ref{sec:background} covers prior work related to our target use case, Section~\ref{sec:methodology} describes our FABRIC setup, 
Section~\ref{sec:results} presents the results of our experiments, and Section~\ref{sec:conclusion} summarizes our findings.

\section{Background and Related Work}
\label{sec:background}

HP-WANs interconnecting facilities that generate and process massive volumes of data have existed for decades, along with established best practices for optimizing data movement. 
These practices include careful tuning of the Data Transfer Nodes (DTNs) deployed at the network edges of Science De-Militarized Zones (DMZs)~\cite{science-dmz}. Prominent examples of such HP-WAN infrastructures, originally developed and operated by the owners of the interconnected data generation and processing facilities, include ESnet, GÉANT, JANET, Effingo, Internet2, CANARIE, and APAN~\cite{hpwan-state-of-art}. Over time, these networks have been opened to a broader community of research organizations, giving rise to a new operational model in which the owner of a data transfer no longer controls the configuration of the entire end-to-end data path.

Transport optimizations for massive data transfers over public networks have been widely studied~\cite{hercules}, and prior work has also evaluated TCP CC choices for Science DMZ DTNs in high-speed wide-area environments~\cite{cubic-vs-bbrv2}. 
However, comparatively little work focuses on the case of large-BDP network paths with guaranteed bandwidth isolation. 
The authors of~\cite{fabric-largebdp}, for example, study the coexistence of TCP flows employing different CC algorithms at a shared network bottleneck under a wide range of BDP conditions and queue management policies. 
In our HP-WAN scenario with per-VC bandwidth isolation, fairness across CC algorithms is not a concern: the owner of the data transfer controls the configuration of all TCP flows within the isolated VC and therefore does not need to multiplex different CC types.

Similarities with our work can be found in~\cite{bbr-evaluation-2017}, including the emulation of a large-BDP VC with complete bandwidth isolation and the use of DPDK for implementing the bottleneck node of the experimental data path. That work, however, focuses on (a) demonstrating the behavior of TCP BBR and its sensitivity to bottleneck buffer sizing, and (b) comparing BBR with TCP CUBIC in scenarios where flows of both types share the VC. 
Our work instead targets public HP-WAN transfers over bandwidth-isolated VCs, where fairness across CC types is not the primary concern and the key objectives are completion-time efficiency and predictability.

Our assumption of a bandwidth-isolated VC between the DTNs involved in a massive transfer is also adopted in~\cite{esnet100gbps}, where the viability of TCP CUBIC and RoCE as HP-WAN transport options is evaluated at 10 and 40\,Gb/s. In that study, RDMA outperforms TCP under data-path conditions that cannot be replicated in a public HP-WAN setting. Moreover, in all experiments involving TCP, the achieved throughput is constrained by the capacity of the sender’s network interface card (NIC). In practical deployments, however, public HP-WAN VCs must be realized through soft isolation over shared resources, with guaranteed bandwidths that are typically well below the NIC capacities of the endpoint DTNs.

With CC algorithms that use the congestion window size (CWND) to regulate the sender data rate, it is well understood that striping the data transfer over parallel TCP connections can maximize the utilization of the bottleneck bandwidth better than a single connection~\cite{gridftp}. 
We share with~\cite{cubic-vs-bbrv2} the choice of traffic generator and load balancing tool (\texttt{iperf3}) for exploration of the benefits of TCP striping. 
The tool implements an ideal load balancer that equalizes the lifetime of the parallel connections independently of their actual throughput.
TCP striping gains relevance in our HP-WAN scenario because complete bandwidth isolation implies the absence of competing traffic at the network bottleneck and thus eliminates all concerns about the unfairness of the striping technique~\cite{hercules}.

Overall, prior work has examined HP-WAN infrastructures and DTN optimization, TCP and RDMA performance in private HP-WAN settings, and TCP coexistence or striping under more general network assumptions. However, these studies do not address the specific case of massive transfers over {\em public\/} HP-WANs with bandwidth-isolated VCs, where users have limited control over the end-to-end data path, fairness is not the primary concern, and the focus is instead on FCT efficiency and predictability. Our work addresses this gap by experimentally comparing representative TCP CC algorithms together with practical ingress data-path configurations.

\section{Experiment Design} \label{sec:methodology}

We conduct experiments on a FABRIC testbed {\em slice\/}~\cite{FABRIC} to assess the performance of CC algorithms under public HP‑WAN conditions, where the user has only partial control over the end‑to‑end data‑path configuration.

\subsection{Fabric Slice Design} \label{sub:slicedesign}

FABRIC supports the provisioning of L2PTP VCs with guaranteed bandwidth between Mellanox ConnectX-6 SmartNICs at remote sites~\cite{fabric-qos}. 
The instantiation of a slice succeeds only if the bandwidth requested for its L2PTP VCs is available on all links of their explicitly assigned paths. The bandwidth is then subtracted from the pool left available for future reservations.
At runtime, a {\em hard policer\/} at the entry point of the VC (the edge-router ingress port that faces the SmartNIC) tightly caps the guaranteed bandwidth of the VC: the policer immediately drops every incoming packet that violates the leaky-bucket profile of the VC without applying color marking~\cite{rfc2697}. The burst size of the leaky-bucket profile is not user-configurable. Relying on the hard enforcement of bandwidth caps at the ingress edge, FABRIC assigns all L2PTP packets to a single {\em expedited forwarding\/}~(EF) queue~\cite{diffserv}. The EF queue holds the highest service priority and is shaped to approximately 80\% of the link capacity; aggregate L2PTP VC reservations are capped at the same fraction of capacity.

\begin{figure}[ht]
    \centering
    \includegraphics[width=0.46\textwidth]{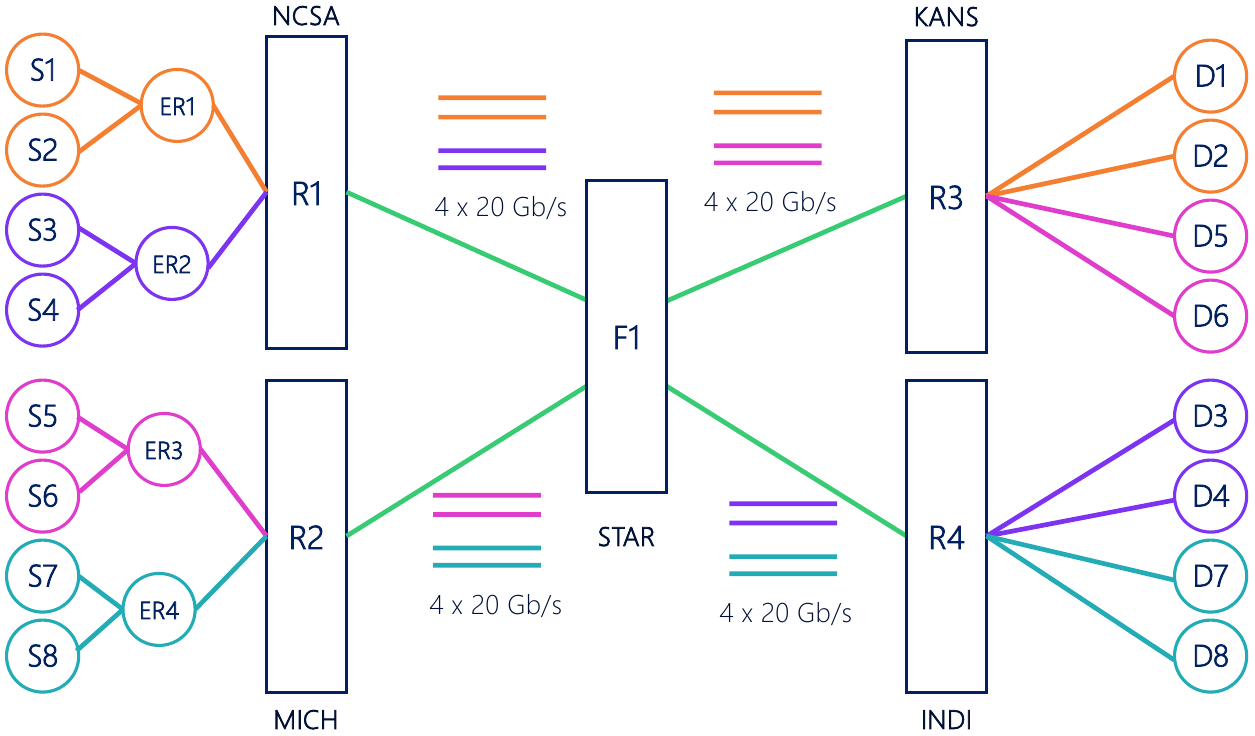}
    \caption{\small{Slice topology; L2PTP VCs terminate on router VMs R1--4.}}
    \label{fig:large-topology}

~\vspace{-3mm}    
\end{figure}

Figure~\ref{fig:large-topology} shows the topology of our slice. Eight sender DTNs at two source sites (in Illinois-based NCSA and in MICHigan) transmit large data files to respective receiver DTNs (in KANSas and INDIana) over high-capacity VCs with non-negligible propagation delay. A central site (STARlight in Illinois) interconnects the source and destination sites. Sender-receiver VCs form pairs with the same base round-trip time (RTT): 14\,ms for S1--2 $\longleftrightarrow$ D1--2, 7\,ms for S3--4 $\longleftrightarrow$ D3--4, 16\,ms for S5--6 $\longleftrightarrow$ D5--6, and 10\,ms for S7--8 $\longleftrightarrow$ D7--8. The topology and VC configuration are intended to produce inter-VC interference at every stage of the data path. Accordingly, we use eight source–destination pairs to create multiple simultaneous VCs, including several with overlapping inter-site paths. This allows us to assess whether VCs sharing the same path experience additional interference.

In experiments with the full topology of Fig.~\ref{fig:large-topology}, the maximum bandwidth reservation of each VC is set to 20\,Gb/s, in order to comply with the 80\,Gb/s reservation limit on the 100\,Gb/s links of FABRIC (R1--4 $\longleftrightarrow$ F1). In system-limit experiments (Section~\ref{sub:systemlimits}), we instead provision a single 80\,Gb/s VC between two sites.

The limited availability of SmartNICs at individual FABRIC sites (typically eight or fewer) makes it difficult to reserve four SmartNICs (one per VC) at each of the four sites with VC terminations. 
To simplify slice creation, we associate all VC terminations at each site with a single SmartNIC, at the cost of increased exposure to cross‑traffic interference.

Accordingly, at each source site we instantiate the DTNs in separate VMs (S1–4 in NCSA and S5–8 in MICH) and group them into pairs, each pair assigned to a distinct host. 
Each source-side DTN host includes an edge-router VM (ER1–4) to optionally insert traffic shaping between the DTNs and the host of the local SmartNIC, on which a router VM (R1–2) forwards traffic to and from the VC endpoints. At the destination sites (KANS and INDI), the DTNs (D1–8) are similarly paired on distinct hosts and communicate directly with the local SmartNIC’s router VM (R3–4); no edge-router VMs are deployed at the destination, as traffic shaping is never required there. All VMs run Ubuntu 22.04.

Packet elevation to the EF priority and enforcement of bandwidth isolation are confined to the data path between the VC endpoints. FABRIC does not provide traffic prioritization between the DTNs and their corresponding VC terminations (i.e., between S1–8 and R1–2, and between D1–8 and R3–4). As a result, in these segments of the network path, data‑transfer packets may experience unpredictable cross‑traffic interference. To assess the effectiveness of traffic shaping in mitigating this interference, we introduce edge‑router (ER) VMs.

The lack of access to performance monitoring counters in the switches along these segments further prevents precise localization of packet losses that are not captured by DTN or router‑VM counters. We refer to such losses as non‑congestive, as they are induced by short‑term traffic burstiness rather than by sustained oversubscription of the reserved bandwidth.

For high-capacity tuning of the DTN VMs, we follow established best practices for 100\,Gb/s TCP~\cite{tcp-100gbps}. 
We increase the socket buffer limits from the Linux default of 16\,MB to 2\,GB, enable jumbo frames, and enable NIC offloading features (TSO, GSO, and GRO) to reduce per-packet processing overhead, while keeping LRO disabled. 
We further apply system-wide Linux \texttt{sysctl} tuning to raise TCP and socket buffer limits, enlarge ingress backlogs, enable MTU probing, disable cached TCP metrics, and set the default queue discipline to Linux~\texttt{tc-fq}. 

On every VM, we disable Linux’s automatic IRQ balancer and manually pin each NIC’s IRQs to a NUMA node to improve packet processing stability. For edge routers (ER1--4) and routers (R1--4), where we evaluate Linux-kernel and DPDK-based packet forwarding, we isolate a set of CPUs for the DPDK applications using \textit{nohz\_full}, \textit{rcu\_nocbs}, and \textit{isolcpus}, while confining OS housekeeping tasks to separate cores via \textit{housekeeping}. 
We also pre-allocate 100 instances of 1\,GB \textit{hugepages} to support high-performance packet processing. 

We pin vCPUs and NUMA-bound memory to the NIC-local socket to minimize cross-socket access and ensure stable high-rate performance. Due to resource constraints when reserving a large number of nodes, CPU pinning and memory binding cannot be applied on a subset of nodes. 
Finally, we enlarge the NIC ring buffers to 8192 descriptors to avoid packet drops and configure the \texttt{tc-fq} queue discipline on sender and receiver interfaces to enable pacing, as recommended in~\cite{l4srepo}.

\begin{table*}[t]
\centering
\setlength{\extrarowheight}{1.2pt} 

\resizebox{\textwidth}{!}{%
\begin{tabular}{|c|c|c|c|}
\hline
 & \textbf{\begin{tabular}[c]{@{}c@{}}ER1--4 \\ (sender-side edge routers)\end{tabular}}
 & \textbf{\begin{tabular}[c]{@{}c@{}}R1--4 \\ (routers at VC endpoints)\end{tabular}}
 & \textbf{Purpose} \\ \hline

\textbf{Linux forwarding}
 & \begin{tabular}[c]{@{}c@{}}Linux kernel forwarding, \\ no shaping\end{tabular}
 & \begin{tabular}[c]{@{}c@{}}Linux kernel forwarding, \\ no shaping\end{tabular}
 & Baseline with the standard Linux datapath \\ \hline

\textbf{DPDK forwarding}
 & \begin{tabular}[c]{@{}c@{}}DPDK fast forwarding, \\ no shaping\end{tabular}
 & \begin{tabular}[c]{@{}c@{}}DPDK fast forwarding, \\ no shaping\end{tabular}
 & Assess the benefit of DPDK forwarding without shaping \\ \hline

\textbf{DPDK shaping}
 & \begin{tabular}[c]{@{}c@{}}DPDK fast forwarding, \\ per-VC shaping\end{tabular}
 & \begin{tabular}[c]{@{}c@{}}DPDK fast forwarding, \\ no shaping\end{tabular}
 & \begin{tabular}[c]{@{}c@{}}Evaluate whether sender-side shaping \\ before VC entry improves performance\end{tabular} \\ \hline
\end{tabular}%
}
\caption{Summary of the three data-path configurations used in the experiments.}
\label{tab:configurations}
\end{table*}

\subsection{Congestion Control Algorithms} \label{sub:congestioncontrol}

We evaluate only congestion control (CC) algorithms designed for high-throughput bulk transfers, excluding approaches that optimize for other objectives such as low packet latency (e.g., L4S \cite{l4srepo}). 
Specifically, we consider: 
(a) TCP CUBIC, the dominant CC algorithm in today’s Internet \cite{census}; 
(b) TCP BBRv1, which is widely deployed in practice \cite{census}; 
and (c) TCP BBRv3, which replaced BBRv2 and is currently used for Google’s internal WAN traffic as well as
google.com and YouTube services~\cite{bbrv3-ietf-120-slides}.

We generate TCP traffic using a multi-threaded version of \texttt{iperf3} (v3.19.1), enabling zerocopy (--zerocopy) and CPU pinning (-A) to reduce CPU overhead and scheduling jitter. 
In experiments with 20\,Gb/s VCs, all eight VCs are concurrently active, each transferring a 1~TB file with FCTs of approximately 8~minutes under ideal conditions.

We run experiments with 1 and 10 parallel TCP flows.
The single-flow case represents a transfer carried by a single TCP connection, while the 10-flow case corresponds to a striped transfer over parallel TCP flows. 
Since our objective is to evaluate the benefits of striping in HP-WAN environments, these two cases provide a direct comparison between unstriped and striped transfers.

Our reference metric is the {\em FCT efficiency\/}, defined as the ratio between the ideal and the measured FCT. The ideal FCT is the file size divided by the bottleneck capacity, adjusted for protocol headers.
We plot the range of FCT efficiency values collected for each VC over 20 independent runs, from minimum to maximum.
Since FABRIC is a public testbed, cross traffic between the DTNs and the VC endpoints cannot be controlled.
Repeating each experiment 20 times, at different times on different days, provides representative coverage of the cross-traffic conditions that the public testbed environment can produce.
For practical purposes, we characterize {\em FCT predictability\/} using the minimum FCT efficiency observed in each 20-run sample, as this value directly quantifies the safety margin that must be added to the ideal FCT when submitting a reservation to ensure completion under adverse conditions.

\subsection{Data-Path Configurations} \label{networksettings}

As an alternative to the Linux kernel, in some experiments we instantiate the packet forwarding function of the routing VMs as a user-space DPDK application.
We test two distinct DPDK applications: one providing plain fast forwarding, and another that additionally supports per-VC queue shaping at the reserved HP-WAN rates~\cite{lightweight-determinism, our_artifacts_github_repo}. 
We focus on Linux kernel forwarding and DPDK because they represent practical data-path configurations in our setting and capture the key contrast between kernel-space forwarding and high-performance user-space packet processing. Our objective is not to provide a broader comparison of I/O frameworks or forwarding engines, but to evaluate the options that are directly deployable in the HP-WAN scenario considered here.

In the fast-forwarding application, a polling thread receives packets on the ingress NIC port, applies MAC/VLAN headers, and forwards packets in bursts to the egress port buffer. In the shaper application, incoming packets are classified into per-sender queues and released to the egress port buffer according to a preconfigured schedule that enforces the desired shaping rates. In both cases, a dedicated TX thread transmits packets from the port buffer, and each routing VM runs two instances of the selected application in parallel to support bidirectional traffic. With Linux kernel forwarding, each routing VM is configured with a multi-queue (\texttt{mq}) root \texttt{qdisc} and per-priority \texttt{pfifo} queues for each traffic direction. Across all routing VM configurations, we provision large buffers (8\,BDP) and enable explicit congestion notification (ECN) with a 1\,BDP marking threshold to minimize local packet loss while preserving congestion visibility.

In the topology of Fig.~\ref{fig:large-topology}, we consider three data-path configurations (summarized in Table~\ref{tab:configurations}):
(a) Linux kernel forwarding on all routers without shaping;
(b) DPDK fast forwarding on all routers;
(c) DPDK shaping on the edge routers (ER1--4) with DPDK fast forwarding on the other routers (R1--4).
Ideally, with (a) and (b) the L2PTP policer should be the sole source of packet loss events, and no losses should occur with (c). 
In practice, additional packet losses are observed due to limited user control over certain segments of the data path.
A VC may lose packets to cross-traffic interference in the shallow buffers of the switches between the DTNs and the VC entry points at both ends of the TCP connection, as well as in the shallow buffers of the HP-WAN path due to collisions with other VCs, even when all VCs are aggressively policed.
Virtualization overheads and jitter may further cause uncounted losses in all VMs.

\section{Experimental Results}
\label{sec:results}

We design our experiments to answer the following question: which combination of data-path configuration and CC algorithm provides the most efficient and predictable massive data transfers over public HP-WAN VCs? We first baseline the system performance limits on a reduced topology, and then evaluate the remaining viable options on the full topology.

\subsection{System Performance Limits}
\label{sub:systemlimits}

We provision a single 80\,Gb/s VC with one sender and one receiver, in line with the logical topology of Fig.~\ref{fig:representative-scenario}, and evaluate different configurations of the traffic manager (instantiated in the R1 router of Fig.~\ref{fig:large-topology}) to assess their ability to sustain high data rates.
For end-to-end transport, we use 10 parallel BBRv1 flows. 
BBRv1 is best suited for identifying system-level bottlenecks because it is relatively insensitive to packet losses and is known to sustain very high sending rates.

\begin{figure}[ht]
    \centering
    \includegraphics[width=0.46\textwidth]{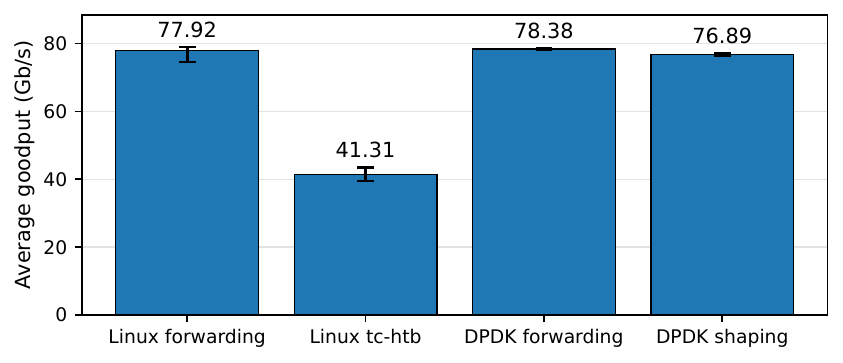}
    \caption{\small{Goodput performance of router VM configuration options.}}
    \label{fig:DPDK-HTB-Linux}
    \vspace{-3mm}
\end{figure}

We evaluate four router configurations: Linux kernel forwarding, with and without traffic shaping (single-queue \texttt{tc-htb}), DPDK fast forwarding, and single-queue DPDK shaping. The shaping rate is set to 80\,Gb/s.
Figure~\ref{fig:DPDK-HTB-Linux} plots the average goodput of five 120-second data transfers for each configuration.
The Linux \texttt{tc-htb} shaper fails to sustain the target goodput due to high processing overhead and is therefore not considered in subsequent experiments.

\begin{figure}[ht]
    \centering
    \includegraphics[width=0.48\textwidth]{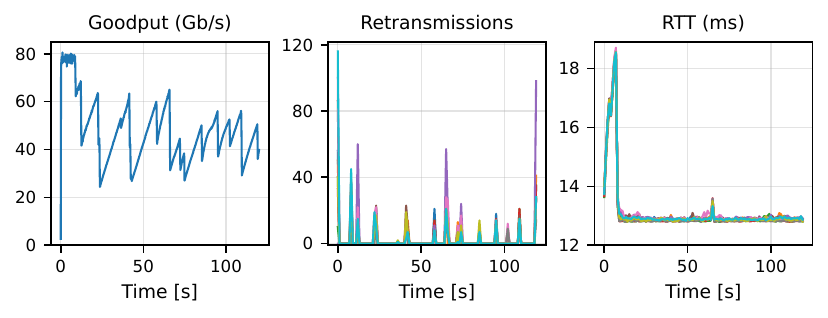}
    \caption{\small{10 TCP CUBIC flows with DPDK shaping at 80\,Gb/s.}}
    \label{fig:cubic}
    \vspace{-3mm}
\end{figure}

We next evaluate the 80\,Gb/s VC using 10 TCP CUBIC flows
with the single-queue DPDK shaper configured at 80\,Gb/s.
While BBRv1 sustains the target rate, TCP CUBIC
does not, as it treats all packet losses as
congestion signals (Fig.~\ref{fig:cubic}). Since this behavior is consistent with known limitations of loss-based CC in such environments, we omit CUBIC from the remaining results and focus on BBRv1 and BBRv3.

\vspace{-2mm}

\begin{figure*}[t]
  \centering
  \captionsetup[subfigure]{skip=5pt}
  \begin{subfigure}{0.98\textwidth}
    \centering
    \includegraphics[width=0.98\linewidth]{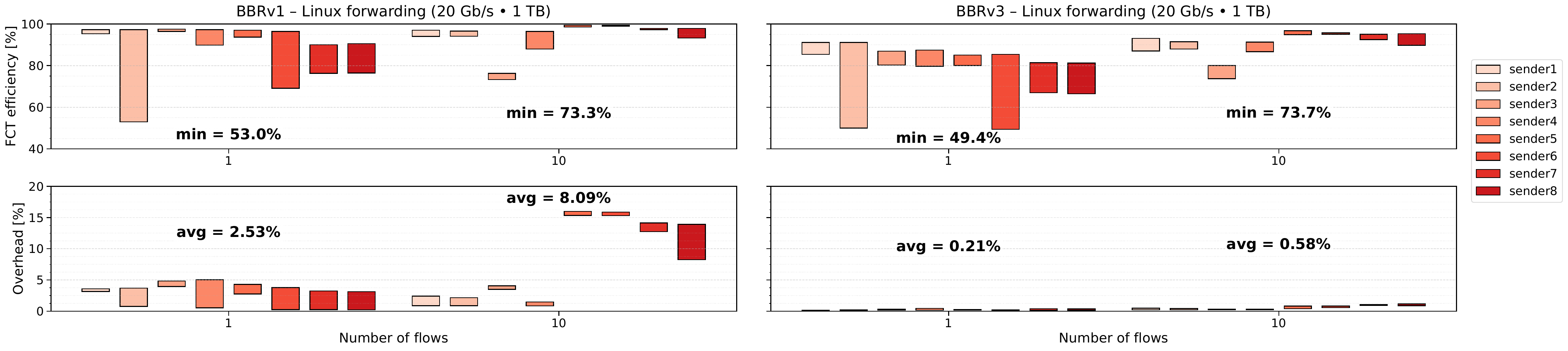}
    \caption{\small{BBRv1 and BBRv3 with Linux forwarding.}}
    ~\vspace{1mm}
    \label{fig:linux-forwarding-20Gbps}
  \end{subfigure}

  \begin{subfigure}{0.98\textwidth}
    \centering
    ~\vspace{1mm}
    \includegraphics[width=0.98\linewidth]{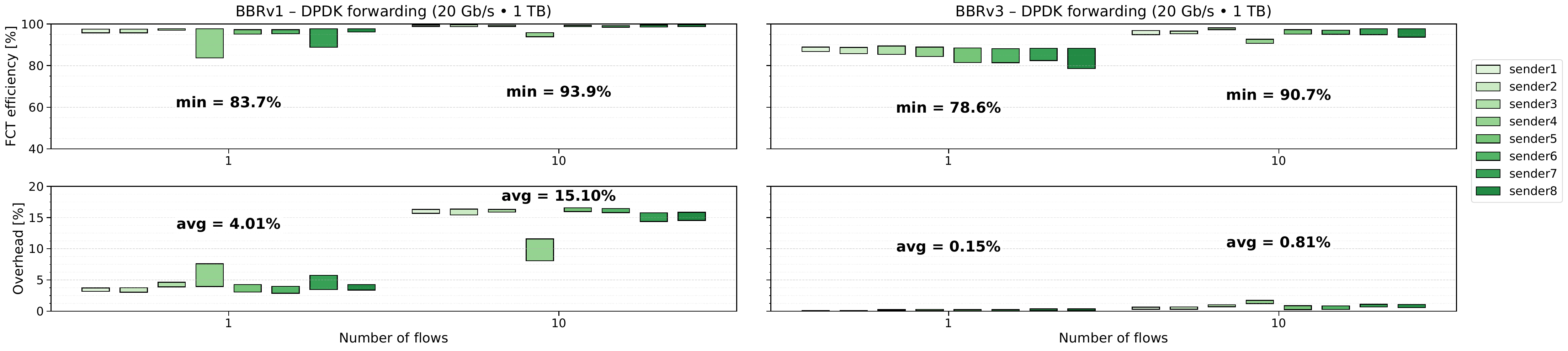}
     \caption{\small{BBRv1 and BBRv3 with DPDK forwarding.}}
     ~\vspace{1.5mm}
    \label{fig:dpdk-forwarding-20Gbps}
  \end{subfigure}

  \begin{subfigure}{0.98\textwidth}
    \centering
    \includegraphics[width=0.98\linewidth]{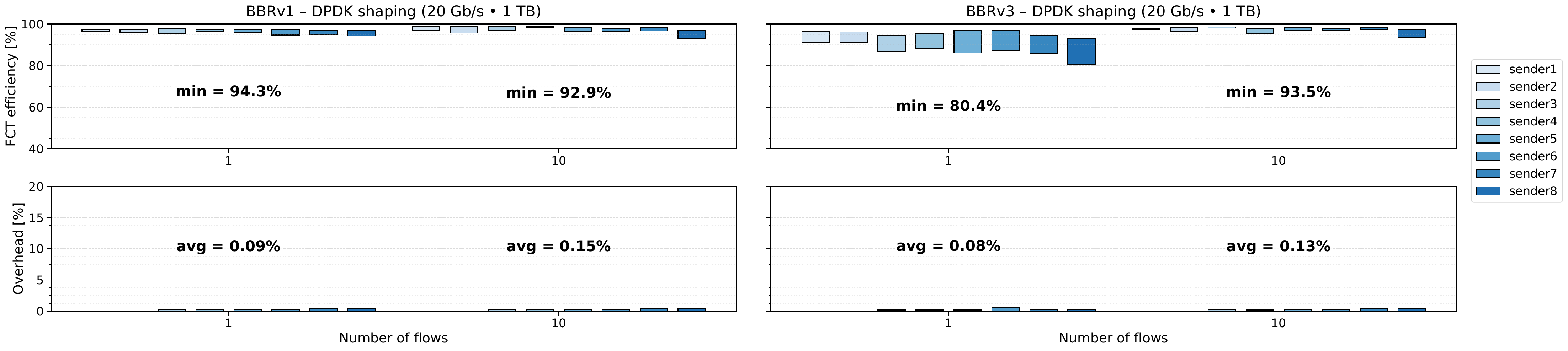}
    \caption{\small{BBRv1 and BBRv3 with DPDK shaping.}}
    \label{fig:dpdk-shaping-20Gbps}
    
  \end{subfigure}
  
  \caption{\small{FCT efficiency and bandwidth overhead for BBRv1 and BBRv3 under different router VM configurations. Bar edges show the min–max values across 20 trials for each sender. Annotations report the minimum FCT efficiency and the average overhead, aggregated over the full set of VCs in each scenario.}}

  \label{fig:main-all}
\end{figure*}

\subsection{Full-Topology Experiments}
\label{sub:fulltopology}

Figure~\ref{fig:main-all} shows the results of our experiments on the full topology of Fig.~\ref{fig:large-topology}. Each bar in the plots refers to one HP-WAN VC. For each combination of data-path configuration and CC algorithm we run experiments with 1 and 10 flows per VC. The overhead plots on the bottom row of each pair report the extra bandwidth consumed by BBR flows between the source VM and the VC entry point due to packet retransmissions, expressed as a percentage of the nominal 20\,Gb/s VC rate. In the DPDK shaping configuration, the per-VC shapers are set to 20\,Gb/s. The main findings are discussed next.

\textbf{DPDK shaping is the best data-path configuration.} 
Per-VC shaping at the edge-router VMs (ER1--4) minimizes traffic burstiness between the TCP senders and the policers, thereby reducing packet losses due to cross-traffic interference, while routers R1--4 operate plain fast forwarding to avoid additional throughput restrictions. 
As a result, DPDK shaping achieves the highest FCT efficiency and predictability, especially with BBRv1, and the lowest worst-case average bandwidth overhead (0.15\%), indicating more accurate estimation of VC bandwidth by the BBR senders. 
In practical terms, minimum FCT efficiencies well above 90\% (BBRv1 with both 1 and 10 flows, and BBRv3 with 10 flows) suggest that a 10\% reservation margin is sufficient. 
Across per-sender results, pairs with similar path delay exhibit comparable performance, confirming that VCs with largely overlapping paths do not interfere with one another.

Linux forwarding relies solely on the hard policers of the
HP-WAN VCs to enforce bandwidth guarantees and results
in the lowest FCT efficiency and predictability among the
evaluated configurations (Fig.~\ref{fig:linux-forwarding-20Gbps}).
We attribute this degradation to the burstiness of high-rate
packet forwarding through the Linux network stack in a
virtualized environment (bare-metal Linux installations are
not feasible in FABRIC). Such burstiness leads to frequent
packet losses in data-path segments outside user control,
including the policers, and critically prevents the BBR senders from accurately estimating the guaranteed bandwidth of the VCs. 

DPDK forwarding reduces inter-packet jitter compared to
Linux forwarding, leading to more stable packet loss
patterns and enabling the BBR sources to better estimate their available bandwidth (Fig.~\ref{fig:dpdk-forwarding-20Gbps}). 
However, the resulting bandwidth overhead is even higher
than with Linux forwarding (15.1\% maximum average vs.
8.09\%). This occurs because loss pattern stabilization is
observed only at the level of the aggregate traffic, while
individual VCs remain bursty, increasing the frequency of
policer-induced packet drops.

\textbf{BBRv1 yields better predictability than BBRv3.}
In our setting with persistent non-congestive packet losses,
BBRv1’s lower sensitivity to loss events yields more stable FCT efficiency across trials and senders than BBRv3, which reduces its sending rate as the frequency of loss events increases. 
Per-VC shaping lowers the loss rate, allowing BBRv3 to
match BBRv1 in the 10-flow case.
This improvement is consistent with the BBRv3 redesign of
\textit{ProbeRTT}: during a \textit{ProbeRTT} episode, BBRv1
reduces the congestion window to four packets for roughly
200\,ms plus one RTT, whereas BBRv3 reduces it to
0.5\,BDP, sustaining a higher sending rate and improving
FCT efficiency when losses are limited.
Finally, in non-shaping configurations BBRv3 exhibits lower
bandwidth overhead than BBRv1 (fewer retransmissions), but
at the cost of lower FCT efficiency.
Together, these results highlight a key trade-off: BBRv1 is
more predictable than BBRv3 under heavy non-congestive losses, while BBRv3 can narrow the gap under data-path configurations that suppress such losses.

\textbf{Shaping below the reserved VC rate provides no benefit.}
With DPDK shaping, we evaluated shaping rates below the
HP-WAN VC reservations (e.g., 15\,Gb/s vs.\ 20\,Gb/s).
We summarize these results without figures for space reasons:
while packet loss rates were reduced, they remained
non-negligible and no meaningful improvement in FCT was
observed. BBRv1 continued to outperform BBRv3, indicating
that robustness to non-congestive losses is a stronger driver of FCT predictability than conservative rate shaping.

\textbf{Non-congestive packet losses are persistent in our FABRIC setup.}
We evaluated several approaches to eliminate non-congestive losses, including deploying per-VC shapers on R1--4 and co-locating shapers with the senders, yet losses persisted.
Although we could not precisely localize the loss points along the path, experiments with co-located senders and shapers suggest that most losses likely occur downstream of the VC egress, close to the TCP receivers. 
These observations indicate that CC with low sensitivity to non-congestive losses is preferable for predictable massive transfers over public HP-WANs. 
However, because BBRv1 can coexist poorly with
loss-based TCP variants, it is best confined to isolated VCs
that do not carry other TCP traffic.

\textbf{Within the tested bandwidth range, single-flow transfers are preferable.}
With BBRv1 and DPDK shaping, single-flow and multi-flow
data transfers exhibit practically equivalent performance.
In our experiments, \texttt{iperf3} emulates an ideal application for packet distribution and reordering across parallel TCP flows. 
Practical applications, however, often introduce
additional coordination and reordering overheads that can
degrade FCT, making single-flow transfers preferable for VC
bandwidths up to at least 20\,Gb/s.

\section{Conclusions}
\label{sec:conclusion}

We studied the completion-time efficiency and predictability
of prominent TCP CC algorithms for massive data transfers
over public HP-WANs. 
Using L2PTP VCs on the FABRIC testbed, we compared three network configurations at the VC ingress (Linux kernel forwarding, DPDK fast forwarding, and DPDK shaping) and found that per-VC shaping substantially improves FCT predictability while minimizing bandwidth overhead from packet retransmissions. 
Combined with per-VC shaping, BBRv1 provides more efficient and predictable file transfers than BBRv3 and CUBIC. Given the challenges posed by the FABRIC data path (especially the
frequent occurrence of non-congestive packet losses), the
pairing of per-VC shaping with BBRv1 emerges as the most
promising option for similar public HP-WANs that support
bandwidth reservations, at least within the 80\,Gb/s range
of our experiments. 
Evaluating additional HP-WAN topologies and data-path options is left for future work.

\bibliographystyle{IEEEtran}
\bibliography{reference}

\end{document}